# Learning Generic Solutions for Multiphase Transport in Porous Media via the Flux Functions Operator

Waleed Diab[a], Omar Chaabi[a], Shayma Alkobaisi[b], Abeeb Awotunde[c], Mohammed Al Kobaisi [a,*]

[a] *Petroleum Engineering Department, Khalifa University of Science and Technology, Abu Dhabi 127788, UAE*
[b] *College of Information Technology, UAE University, Al Ain 15551, UAE*
[c] *College of Petroleum Engineering and Geosciences, KFUPM, Dhahran, Saudi Arabia*

**ABSTRACT**

Traditional numerical schemes for simulating fluid flow and transport in porous media can be computationally expensive. Advances in machine learning for scientific computing have the potential to help speed up the simulation time in many scientific and engineering fields. DeepONet has recently emerged as a powerful tool for accelerating the solution of partial differential equations (PDEs) by learning operators (mapping between function spaces) of PDEs. In this work, we learn the mapping between the space of flux functions of the Buckley-Leverett PDE and the space of solutions (saturations). We use Physics-Informed DeepONets (PI-DeepONets) to achieve this mapping without any paired input-output observations, except for a set of given initial or boundary conditions; ergo, eliminating the expensive data generation process. By leveraging the underlying physical laws via soft penalty constraints during model training, in a manner similar to Physics-Informed Neural Networks (PINNs), and a unique deep neural network architecture, the proposed PI-DeepONet model can predict the solution accurately given any type of flux function (concave, convex, or non-convex) while achieving up to four orders of magnitude improvements in speed over traditional numerical solvers. Moreover, the trained PI-DeepONet model demonstrates excellent generalization qualities, rendering it a promising tool for accelerating the solution of transport problems in porous media.

## 1. Introduction

Hitherto, numerical schemes have been the only viable mean for modeling complex physical processes across a wide spectrum of engineering disciplines. Recently, however, machine learning techniques have proven to be potent tools for tackling complex problems across various scientific domains and may very well be a legitimate contender in the near future. They are employed in a broad spectrum of areas, ranging from computational fluid dynamics [1], [2] and molecular simulations [3] to climate modeling [4], [5] and high-energy physics [6]–[8]. The advent of Physics-Informed Neural Networks [9] (PINNs) and operator learning [10]–[15] algorithms grounded in deep learning has paved the way for innovative research in scientific computing, especially concerning physics-driven problems. Many recent works [16]–[23], rely on the advent of these innovative algorithms to solve PDEs with little or no data, where the training is guided by the PDE itself through automatic differentiation [24]. These works underscore the significant role of integrating established physics, conservation laws, and constitutive rules to guide the training of neural networks. Such an approach often yields superior generalization and accuracy. Moreover, incorporating physical principles enhances the explainability of deep learning models, countering the common perception of these models as 'black boxes'.

______________________
* Corresponding author at: PE Department, Khalifa University of Science and Technology, Abu Dhabi 127788, UAE
  *E-mail address*: mohammed.alkobaisi@ku.ac.ae (M. Al Kobaisi)

Accurate and efficient modeling of fluid transport in porous media is a critical area of research that carries significant implications for fields such as hydrocarbon recovery [25], [26], hydrogen storage and $CO_2$ sequestration [27]–[29], groundwater remediation [30], [31], and soil sciences [32], [33]. Accurate modeling and forecasting of fluid flow, solute transport, and reactive processes in porous media are vital for reliable predictions of flow, better understanding of reservoirs, and reducing uncertainty in these systems [34]. Recent studies [23], [35]–[40] focused on flow and transport in porous media have been delving into diverse physics-informed neural network architectures to attain these objectives. However, these studies share the same drawback of traditional numerical solvers, as they are generally limited in their flexibility in terms of parameters such as initial and boundary conditions, relative permeability, absolute permeability and porosity maps, and source terms. In other words, any alteration in any of the parameters requires a retraining of the neural network [41] or transfer learning [42]. However, recent advances in operator learning algorithms, such as [10]–[12], promise to address this problem.

The role of modeling extends beyond just simulating physical processes; it also involves leveraging these models to make informed decisions, which often entails running thousands of simulations to account for a wide range of scenarios [43]. Despite the widespread use of traditional numerical methods like finite element and finite volume methods to tackle these large-scale problems [26], [34], they often demand considerable computational resources, particularly when dealing with complex geometries, highly heterogeneous media, and nonlinear processes [44]. This led to the development of reduced order models (ROMs) to manage the computational burden of full-scale numerical models [45]–[48]. ROMs capture the essential dynamics or behavior of complex systems, while significantly reducing the dimensionality and computational cost at the expense of accuracy and generalizability. Our ultimate goal is to develop a method, leveraging machine learning, to lessen the computational burden of stochastic simulations while maintaining the accuracy of full-scale simulation.

Solving a Partial Differential Equation (PDE) across a range of parameters can be mathematically framed as solving a parametric PDE [49], where these parameters are allowed to vary over a certain range of interest. In essence, it is equivalent to identifying the solution operator that maps the input parameter space to the corresponding latent solution space governed by the PDE. In this paper, we leverage neural operator learning algorithms to solve parametric transport problems in porous media. By incorporating physical knowledge and operator learning, Physics-Informed Deep Operator Networks (PI-DeepONets) offer a potent and adaptable approach to solving complex problems while maintaining computational efficiency [50]. It is important to highlight that PI-DeepONets do not require paired input-output observations, a feature we capitalize on in this work to bypass the need for costly numerical simulations.

We will demonstrate the effectiveness of our proposed PI-DeepONet framework through a series of case studies, highlighting its capabilities in handling various types of flux functions. We then proceed to train a model that is capable of handling all types of flux functions concurrently. Furthermore, we will compare the performance of our approach with traditional numerical methods showcasing the advantages of PI-DeepONets in the context of transport in porous media. The comparison is made based on accuracy and efficiency. To the best of our knowledge, this work represents the first application of data-free, operator learning and physics-informed operator learning to transport in porous media problems. In addition, this is the first application of operator learning algorithms to a hyperbolic PDE where the mapping is performed from the space of flux functions to the space of solutions with or without paired input-output observations.

The paper proceeds as follows. In section 2, we present the two-phase governing equation for transport in porous media. Section 3 provides an overview of physics-informed DeepONets. In section 4, we solve the parametric transport problem in one dimension using the PI-DeepONet. Three cases are provided, one for each type of flux function highlighting PI-DeepONet's capability in capturing complex physics such as waves, shocks, and rarefactions. A fourth case is then presented where we train a PI-DeepONet on the three types of flux functions concurrently. The trained model can capture all possible kinds of solution behaviors in one single model in an unsupervised manner, i.e. without any paired input-output observations. Finally, section 5 provides a summary of our main findings, conclusions, and future research directions.

## 2. Transport in Porous Media

In this work, we employ Physics-Informed Deep Operator Network (PI-DeepONet) [50] to solve parametric transport problems in porous media. These problems involve various types of nonlinear flux functions that lead to nonlinear transport problems which can be challenging to solve numerically [26] as the solution manifests mixed waves. A two-phase, immiscible fluid transport in porous media can be described by a nonlinear hyperbolic PDE, in one dimension, as follows:

$$\frac{\partial S_w}{\partial t_D} + \frac{df(S_w, M)}{dx_D} = 0, \tag{1}$$

subject to uniform initial and boundary conditions:

$$\begin{aligned} S_w(x_D, 0) &= S_{wc}, & \forall x_D \\ S_w(x_D, t_D) &= S_{\alpha r}, & x_D = 0 \text{ and } t_D > 0. \end{aligned} \tag{2}$$

The saturation, denoted by $S = S(x, t)$, represents the dependent variable of interest. As a dimensionless property, the physical range of $S_w$ lies between 0 and 1. Equations (1) and (2) are dimensionless, where $t_D$ and $x_D$ are dimensionless time and space, respectively. The flux function, $f(S, M)$, is a nonlinear function also referred to as the fractional flow of water and is defined as:

$$f_w = \frac{1}{1 + M}. \tag{3}$$

The water fractional flow function (3) can be of three types; concave, convex, or nonconvex [38], depending on $M$ which is a physical parameter known as the mobility ratio and defined as follows:

$$M = \frac{\lambda_\alpha}{\lambda_w} = \frac{k_{r\alpha} \mu_w}{k_{rw} \mu_\alpha}, \tag{4}$$

where $\lambda$ is the phase ($\alpha$) mobility, $k_r$ is the relative permeability, and $\mu$ is the fluid viscosity. The type of flux function $f(S, M)$ is determined by a combination of parameters that make-up the relative permeability model. The Honarpour empirical model [51] is commonly used to capture the relationship between phases flowing through a porous medium and is given by:

$$k_r = k_r^* \left( \frac{S_w - S_{wc}}{1 - S_{wc} - S_{\alpha r}} \right)^n, \qquad (5)$$

where $k_r^*$ is the relative permeability endpoint for each phase, $S_{wc}$ and $S_{\alpha r}$ are the connate and residual saturations of water and the second phase ($\alpha$), respectively, and $n$ is the relative permeability exponent. The PDE under investigation here is referred to in the petroleum literature as the classical Buckley-Leverett problem [52] which describes the displacement of one incompressible, immiscible fluid by another in a porous medium, under the assumption of one-dimensional, horizontal flow. This problem is often applied to analyze the displacement of oil by water (waterflooding) in oil reservoirs. For a full description of the problem see [25], [34], [53], [54], and for a mathematical perspective see [55].

Equation (1) represents a hyperbolic Partial Differential Equation (PDE) which has been previously addressed using the PINN approach [35], [36], [38], [56], [57]. As demonstrated in [38], the PINN approach encounters difficulties in solving the hyperbolic PDE in scenarios where shocks are present in the case of convex and nonconvex flux functions. The solution to this issue has been found in the addition of a diffusion term, see [38], that smoothens the shock-front, thereby enabling PINN to effectively solve the parabolic equation which takes the following form:

$$\frac{\partial S_w}{\partial t_D} + \frac{df(S_w)}{dx_D} - \epsilon \frac{\partial^2 S_w}{\partial x_D^2} = 0, \qquad (6)$$

where $\epsilon \geq 2.5 \times 10^{-3}$. The failure of PINN in solving the hyperbolic Buckley-Leverett problem where a shock is present can be largely attributed to the discontinuity in the gradient of the hyperbolic PDE. This results in a failed training process unless a diffusion term is incorporated. In this work, we will be solving the PDE in (6) instead of (1) as we know that PINN fails in solving (1) which we found to be also true in PI-DeepONets. In non-shock problems, both PINN and PI-DeepONets work well without the addition of a diffusion term.

The primary objective of this study, however, is to learn the solution operator, which maps any type of flux function $f_w$ to the corresponding PDE solutions $S_w$. This is achieved in a self-supervised manner using a physics-informed DeepONet, requiring no labeled data except for the initial and boundary conditions. Typically, flux functions are produced by generating relative permeability curves. This process can involve up to four parameters per phase, totaling eight parameters: phase viscosities ($\mu_w$ and $\mu_\alpha$), relative permeability endpoints ($k_{rw}^*$ and $k_{r\alpha}^*$), relative permeability exponents ($n_w$ and $n_\alpha$), and residual saturations ($S_{wc}$ and $S_{\alpha r}$). The last two parameters, $S_{wc}$ and $S_{\alpha r}$, correlate directly with the boundary and initial conditions, respectively. This implies that to include these parameters as variables in the flux functions, the boundary, and initial conditions must vary, which necessitates the learning of multiple operators simultaneously. This is currently beyond the model's capabilities and is a topic for future research. However, by varying the other six parameters, we can produce any type of flux function, enabling the solution of the parametric PDE (1) in more realistic scenarios.

## 3. Learning Operators with Physics Informed DeepONets

Our focus is confined to physics-driven machine learning algorithms, as they provide models grounded in physics that eliminate the need for expensive numerical simulations while simultaneously facilitating data assimilation [2]. The most notable algorithm in this context is the Physics-Informed Neural Network (PINN) [9], renowned for its integration of physics into deep learning models [2]. By imposing governing equations and physics laws as soft-penalty constraints during training, PINNs can learn the inherent dynamics of a physical system and simultaneously comply with governing physics [9]. This method of integrating physics principles into the machine learning framework substantially enhances the model's generalization capabilities, lessens the dependency on large data sets for training, and mitigates the risk of overfitting, a common problem with deep learning models [58]. PINNs drawback is that it is trained for specific initial and boundary conditions, flux function, source term, etc. where any alteration in any of these parameters requires a retraining of the neural network rendering PINNs expensive for inference [41]. To reduce this computational expense, some suggest employing transfer learning [42]. Alternatively, physics-informed operator learning algorithms such as Physics-Informed Deep Operator Networks (PI-DeepONets) offer an alternate approach [50]. PI-DeepONets retain many of the advantages of PINNs while overcoming some of the aforementioned limitations. These networks, engineered specifically to learn and approximate linear and nonlinear operators, treat the operator as an unknown function and approximate it using a deep neural network [11]. This strategy is particularly beneficial in scientific computing, where obtaining the underlying operators analytically or through computationally-intensive simulations can be challenging. In this section, we will present an overview of the physics-informed DeepONet architecture for learning the solution operator of parametric PDEs.

Physics-Informed deep operator networks are an emerging deep learning framework aimed at learning abstract nonlinear operators between infinite-dimensional functional spaces from a finite collection of observed inputs that are governed by partial differential equations and other physics which may or may not include observed outputs (solutions). It draws inspiration and validation from the universal approximation theorem for operators, which provides the theoretical basis for this new paradigm in deep learning directed at learning solution operators for parametric partial differential equations (PDEs) [11].

Consider the following general parametric PDE:

$$\mathcal{N}(\boldsymbol{u}, \boldsymbol{s}) = 0, \tag{7}$$

subject to the following boundary conditions:

$$\mathcal{B}(\boldsymbol{u}, \boldsymbol{s}) = 0, \tag{8}$$

where $\mathcal{N}: \mathcal{U} \times \mathcal{S} \to \mathcal{V}$ is a linear or nonlinear differential operator that represents a mapping between the abstract function spaces $(\mathcal{U}, \mathcal{V}, \mathcal{S})$, $\boldsymbol{u} \in \mathcal{U}$ denotes the input functions, and $\boldsymbol{s} \in \mathcal{S}$ denotes the unknown latent functions that are governed by the PDE. This system is subject to the boundary conditions operator $\mathcal{B}[\cdot]$ that enforces any Dirichlet, Neumann, Robin, or periodic boundary conditions. We make no distinction in this formulation between time $t$ and space $\boldsymbol{x}$, and $\Omega$ covers the full spatiotemporal domain.

We aim to learn the PDE solution operator $G: \mathcal{U} \to \mathcal{S}$ as

$$G(\boldsymbol{u}) = \boldsymbol{s}(\boldsymbol{u}), \tag{9}$$

assuming that there exists a unique solution $\boldsymbol{s} = \boldsymbol{s}(\boldsymbol{u}) \in \mathcal{U}$ to the PDE given appropriate initial and boundary conditions for any $\boldsymbol{u} \in \mathcal{U}$. The operator $G$ is represented by DeepONet which we refer to as $G_\theta$, and $\theta$ denotes the set of all trainable parameters of the DeepONet. The DeepONet has a unique architecture as it is composed of two separate neural networks referred to as the "branch" and "trunk" networks, respectively, as shown in Figure 1. The branch network takes a function $\boldsymbol{u}$ as input where $[u(s_1), u(s_2), \ldots, u(s_m)]$ represents a function $\boldsymbol{u} \in \mathcal{U}$ evaluated at a collection of fixed saturations $\{s_i\}_{i=1}^m$, and learns features embedding $[b_1, b_2, \ldots, b_q]^T \in \mathbb{R}^q$ as output. The fixed saturations $\{s_i\}_{i=1}^m$ are referred to as sensor locations. The trunk network takes the continuous coordinates of the spatiotemporal domain $\boldsymbol{y}$ as inputs, and learns features embedding $[t_1, t_2, \ldots, t_q]^T \in \mathbb{R}^q$. The final DeepONet output is the inner product of the outputs of branch and trunk networks. Mathematically, one can express the DeepONet $G_\theta$ representation of a function $\boldsymbol{u}$ evaluated at $\boldsymbol{y}$ for vector-valued functions as

$$G_\theta^{(i)}(\boldsymbol{u})(\boldsymbol{y}) = \sum_{k=q_{i-1}+1}^{q_i} b_k\big(\boldsymbol{u}(s_1), \boldsymbol{u}(s_2), \ldots, \boldsymbol{u}(s_m)\big) t_k(\boldsymbol{y}), \quad i = 1, \ldots, n, \tag{10}$$

where $0 = q_0 < q_1 < \cdots < q_n = q$.

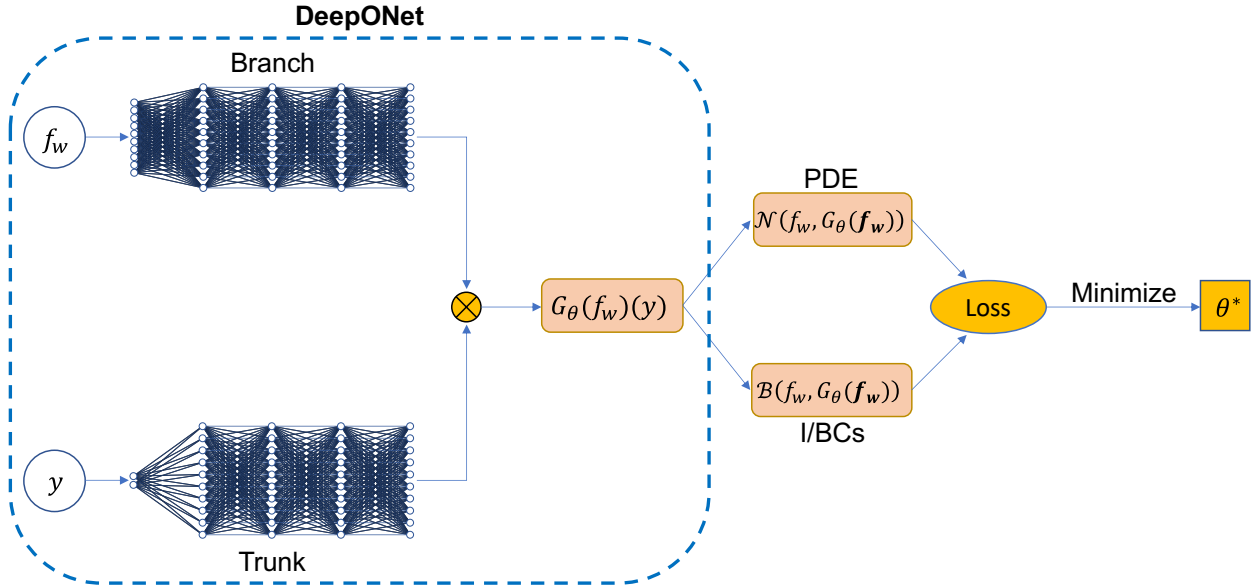

**Fig. 1:** The DeepONet consists of a branch network and a trunk network. The trunk network processes spatial information, while the branch network processes the functional relationships among input functions. A continuous and differentiable output is then obtained through an inner product of the feature vectors ($b_k$ and $t_k$) from the branch and trunk networks. By leveraging automatic differentiation we can construct the PDE from the output of the DeepONet and use it as a soft-penalty constraint to guide the training of the model.

To this end, the DeepONet is not yet informed of the PDE; this is achieved through automatic differentiation to bias the loss function similar to Physics-Informed Neural Networks (PINNs). The loss function for PI-DeepONets is as follows

$$\mathcal{L}(\theta) = \mathcal{L}_{BC}(\theta) + \mathcal{L}_{PDE}(\theta), \tag{11}$$

where

$$\mathcal{L}_{BC}(\theta) = \frac{1}{N}\sum_{i=1}^{N} \mathcal{L}_{BC}(\boldsymbol{u}^{(i)}, \boldsymbol{\theta}) = \frac{1}{NP}\sum_{i=1}^{N}\sum_{j=1}^{P} \left| \mathcal{B}\left(u^{(i)}\left(\boldsymbol{s}_j^{(i)}\right), G_\theta\big(u^{(i)}\big)\big(\boldsymbol{y}_j^{(i)}\big)\right)\right|^2, \tag{12}$$

and

$$\mathcal{L}_{PDE}(\theta) = \frac{1}{N}\sum_{i=1}^{N}\mathcal{L}_{PDE}(u^{(i)},\theta) = \frac{1}{NQ}\sum_{i=1}^{N}\sum_{j=1}^{Q}\left|\mathcal{N}\left(u^{(i)}\left(s_{r,j}^{(i)}\right),G_{\theta}(u^{(i)})\left(y_{r,j}^{(i)}\right)\right)\right|^{2}. \qquad (13)$$

To enforce the set of boundary and initial conditions described in equation (8), $N$ input samples $u^{(i)}$ evaluated at $P$ locations in the domain of $G(u)$ are used, and for each input pair $u^{(i)}$ and $y_j^{(i)}$, $\{s_j\}_{j=1}^{P}$ are randomly sampled from the domain of $u$ and $\{y_j\}_{j=1}^{P}$ are randomly sampled from the boundary of $G(u)$. In addition, for enforcing the set of parametric PDE constraints described in equation (11), $N$ input samples $u^{(i)}$ evaluated at $Q$ location (referred to as collocation points) for evaluating the PDE residual and $\{s_{r,j}\}_{j=1}^{Q}$ and $\{y_{r,j}\}_{j=1}^{Q}$ are collocation points randomly sampled from the domain of $u$ and $G(u)$, respectively, are used. In [50] it was shown that PI-DeepONets can learn the solution operator of parametric PDEs in a self-supervised manner, i.e. without any output observations. This is substantiated in our findings.

In this study, all networks utilize the tanh activation function for all hidden layers. Furthermore, we employ the sigmoid activation function for all the inner products of the branch and trunk networks' outputs, as this strategy aims to ensure a physically consistent and monotone solution [59]. For cases 4.3 and 4.4, where the model has to learn more complex features, the modified DeepONet architecture [60] was used and demonstrated considerable improvement in the overall results. This particular architecture features two encoders applied to the inputs of the trunk and branch networks, which enhances the networks' ability to handle nonlinearities [60]. Comprehensive details of this architecture are beyond the scope of this paper; for more information see Appendix A, and readers are referred to the original work in [60].

## 4. Results

To evaluate the capabilities of the physics-informed DeepONet (PI-DeepONet) framework for solving nonlinear transport problems in porous media, we begin by recasting equations 1 and 2 in the form of a nonlinear parabolic transport equation:

$$\frac{\partial S_w}{\partial t} + \frac{\partial f(S_w)}{\partial x} - \epsilon\frac{\partial^2 S_w}{\partial t^2} = 0, \qquad (14)$$

with the boundary conditions

$$\begin{aligned} S_w(x,0) &= S_{wc}, & x \in [0,1] \\ S_w(0,t) &= S_{\alpha r}, & t > 0. \end{aligned} \qquad (15)$$

We aim to learn the solution operator for mapping $f(S_w)$ to the corresponding PDE solutions $S(x,t)$ using a physics-informed DeepONet. We approximate the operator with a physics-informed DeepONet architecture $G_\theta$, where the branch and trunk networks are two separate neural networks. For a given input function $f^{(i)}(S_w)$, we define the corresponding PDE residual as

$$\mathcal{R}_\theta(f_w)(x,t) = \frac{\partial G_\theta(f_w)(x,t)}{\partial t} + \frac{\partial f_w}{\partial G_\theta}\frac{\partial G_\theta(f_w)(x,t)}{\partial x} - \epsilon\frac{\partial^2 G_\theta(f_w)(x,t)}{\partial t^2}, \qquad (16)$$

where $[f_w(s_1), f_w(s_2), \ldots, f_w(s_m)]$ represents the input functions, and $\{s_i\}_{i=0}^{m}$ is a collection of equispaced sensors in $[s_{wc}, 1 - s_{\alpha r}]$. The parameters $(\theta^*)$ of the physics-informed DeepONet can be trained by minimizing the loss function

$$\mathcal{L}(\theta) = \lambda_{IC} \mathcal{L}_{IC}(\theta) + \lambda_{BC}\mathcal{L}_{BC}(\theta) + \lambda_{PDE}\mathcal{L}_{PDE}(\theta) \tag{17}$$

$$= \lambda_{IC} \frac{1}{NP} \sum_{i=1}^{N} \sum_{j=1}^{P} \left|G_\theta(u^{(i)})(x_{ic,j}^{(i)}, 0) - s_{wc}^{(i)}\right|^2 + \lambda_{BC} \frac{1}{NP} \sum_{i=1}^{N} \sum_{j=1}^{P} \left|G_\theta(u^{(i)})(0, t_{bc,j}^{(i)}) - s_{\alpha r}^{(i)}\right|^2$$

$$+ \lambda_{PDE} \frac{1}{NQ} \sum_{i=1}^{N} \sum_{j=1}^{Q} \left|\mathcal{R}_\theta^{(i)}(x_{r,j}^{(i)}, t_{r,j}^{(i)})\right|^2, \tag{18}$$

where $\lambda_{IC}, \lambda_{BC}$, and $\lambda_{PDE}$ are manually set weights to balance the loss contribution of each component and to help improve the training process [60]–[62].

Here, for each $f_w^{(i)}$, $\{y\}_{j=1}^{P} = \left\{\left(x_{u,j}^{(i)}, t_{u,j}^{(i)}\right)\right\}_{j=1}^{P}$ are uniformly sampled points from the boundary of $[0,1] \times [0,1]$, while $\left\{\left(x_{r,j}^{(i)}, t_{r,j}^{(i)}\right)\right\}_{j=1}^{Q}$ is the set of collocation points uniformly sampled in $[0,1] \times [0,1]$. We train all the physics-informed DeepONets in this work by minimizing the loss function (17) using the Adam optimizer and a batch size of 10,000.

We should emphasize that for all the cases discussed in this study, the model solely receives data in the form of flux functions $f_w(S)$ and initial and boundary conditions expressed as residual saturations. This approach is entirely data-free, requiring no information about the distribution of the solution $S(x, t)$ beyond that provided at the boundaries.

It is also important to acknowledge the inherent connection between the initial and boundary conditions ($s_{wc}$ and $s_{\alpha r}$) and the flux function $f_w(S)$. The flux function is a continuous function evaluated over the saturation range $[s_{wc}, 1 - s_{\alpha r}]$. This relationship limits the model's ability to learn flux function with varying residual saturations as that would entail learning multiple operators simultaneously.

To evaluate the solutions obtained from various trained models, we solve the problem analytically using the method of characteristics (MOC) and via a finite difference scheme. Different types of flux functions result in distinct analytical solutions [38], rendering it challenging to obtain a universal analytical solution that can be used to benchmark the PI-DeepONet solutions. Therefore, we also use an explicit upwind finite difference scheme to compare with the PI-DeepONet. Nonetheless, we will present the analytical solution for comparison where appropriate. All solutions (PI-DeepONet, MOC, Numerical) are evaluated on a $200 \times 200$ equispaced (spatiotemporal) grid to mitigate grid artifacts. The finite difference solution, however, is computed on a $200 \times 1000$ grid to achieve a high-fidelity solution and to abide by the Courant–Friedrichs–Lewy (CFL) condition but the solution in time is saved in increments of 5 to yield a $200 \times 200$ solution. A grid sensitivity study is provided in Appendix B.

Furthermore, we evaluate the PI-DeepONet solution quality both qualitatively and quantitatively. Qualitative evaluation is conducted by overlaying the PI-DeepONet solution on the numerical/analytical solution at three equidistant time steps. For quantitative assessment, we generate 100 new input functions not present in the training set ($N_{test} = 100$) and report the mean and standard deviation of the relative L2 prediction errors between the PI-DeepONet and the finite-difference solution.

## 4.1 Concave Flux Function

If the relative phase permeabilities, $k_r(S)$, are linear functions of saturation, and if we denote the ratio of the phase viscosities as $M = \frac{\mu_o}{\mu_w}$, the corresponding flux function, $f_w$, can be reduced to the following:

$$f_w(S_w) = \frac{S_w}{S_w + \frac{1-S_w}{M}}. \tag{19}$$

This flux function is a linear function of water saturation, and it is concave for $M > 1$. In this case, the solution is only comprised of rarefaction (spreading) waves. Therefore, for our first test case, we approximate the operator $G_\theta$ mapping the space of concave flux functions to the space of solutions of the Buckley-Leverett equation. We sample 1000 unique mobility ratios from a uniform random distribution such that $1 \leq M \leq 10$. Using these mobilities we generate $N_{train} = 1000$ concave flux functions $f_{w_{concave}}^{(i)} = [f_w^{(i)}(s_1), f_w^{(i)}(s_2), \ldots, f_w^{(i)}(s_m)]$ evaluated at $m = 200$ sensor locations, and we take $P = Q = m$ (Figure 2). The weights in equation (17) are all set to one ($\lambda_{IC} = \lambda_{BC} = \lambda_{PDE} = 1$). The vanilla physics-informed DeepONet which consists of branch and trunk networks with two separate 4-layer fully-connected neural networks with 100 neurons per hidden layer is trained for 300,000 iterations using the Adam optimizer with an exponentially decaying learning rate of 0.95 every 1000 steps and an initial learning rate of $1 \times 10^{-3}$.

Figure 3 shows a comparison between the predicted, the analytical, and the finite difference solutions for three test samples at three different time steps along with their associated flux functions. The mean and standard deviation of the relative $L^2$ error for $N_{test} = 100$ test samples are $8.53 \times 10^{-3} \mp 1.84 \times 10^{-3}$, respectively. The trained model captures the physical behavior of the rarefaction waves well, including the speed and size of these waves, although smoothing around sharp edges is sometimes observed as in Figure 3(c). The relative $L^2$ errors of the cases shown in Figure 3 are $4.87 \times 10^{-3}$, $1.99 \times 10^{-3}$, and $4.06 \times 10^{-3}$, for cases a, b, and c, respectively. The trained PI-DeepONet model achieves more than four orders of magnitude improvement in speed compared to the numerical solver running on the same GPU in inference. Here, it is important to keep in mind that the PI-DeepONet model predictions are made over the entire grid ($200 \times 200$) in one shot, while the numerical model solves the problem one-time step at a time. In addition, the numerical model is solved on a ($200 \times 1000$) grid to achieve comparable accuracy, this is true for all the subsequent cases in this paper.

The trained physics-informed DeepONet network shows remarkable accuracy and generalizes well to unseen test functions. It also generalizes well to mobility ratios outside the training range as shown in Figure 4, albeit the solution is less accurate with a relative $L^2$ error of $1.13 \times 10^{-2}$. The solution quality degrades the further the mobility ratio is from the original training space ($1 \leq M \leq 10$). To evaluate the model generalizability to mobility ratios outside the training domain we generate 100 mobility ratios from a uniform random distribution such that $10 \leq M \leq 20$ and generate a new test set. The mean and standard deviation of the relative $L^2$ error of 100 test samples sampled from outside the training range is an order of magnitude lower than that of the original training set range; $8.64 \times 10^{-2} \mp 4.30 \times 10^{-2}$, respectively. Not surprisingly, the further the mobility ratio from the training range the less accurate the solution becomes.

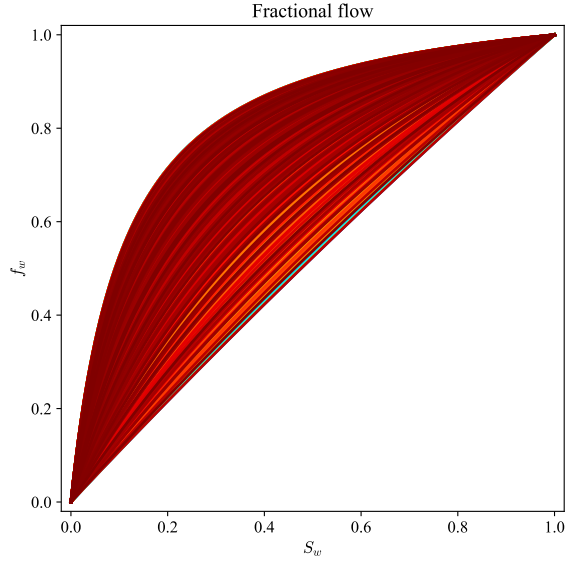

**Fig. 2:** Concave flux functions comprising the training data set ($N_{train} = 1000$).

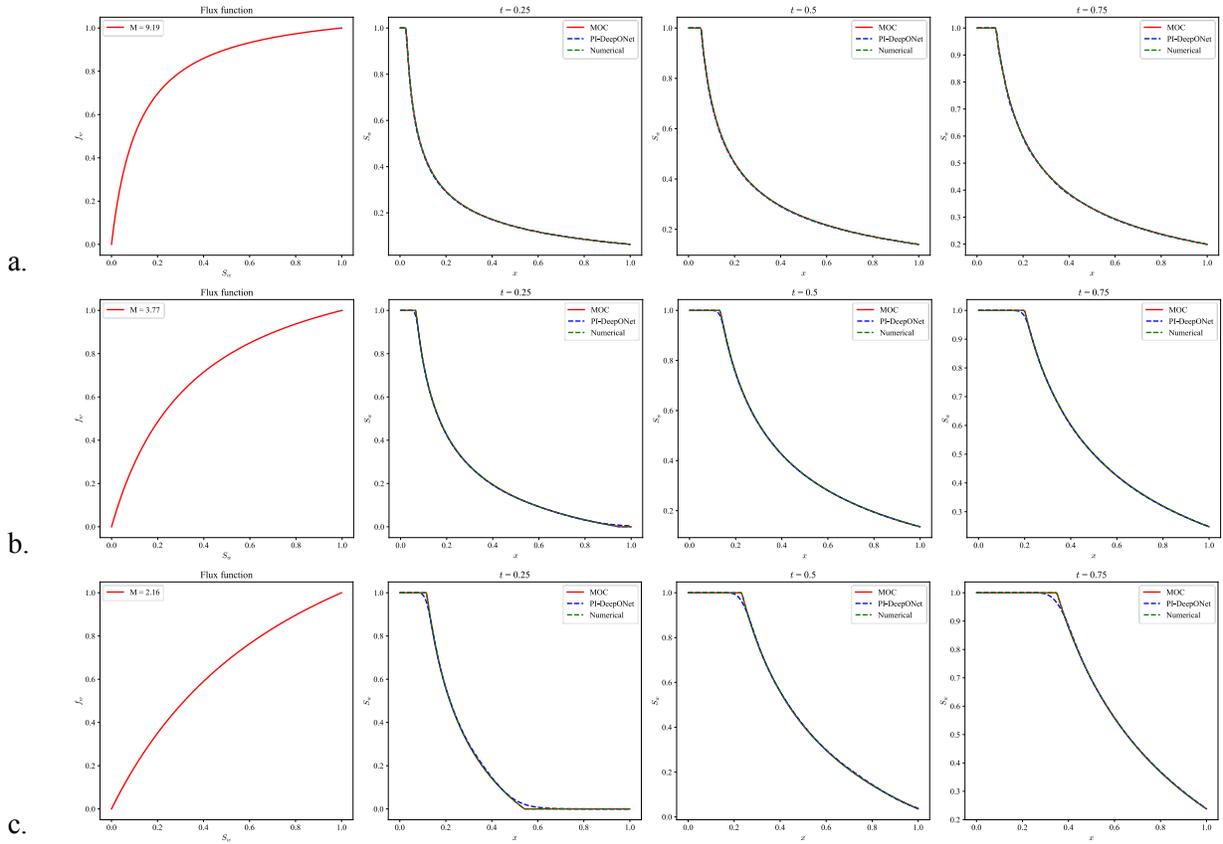

**Fig. 3:** Data-free solution of parametric Buckley-Leverett problem with three different concave flux functions using PI-DeepONets. a. shows the solution profiles at three different time steps for a concave flux function generated using a mobility ratio of 9.19. In the second and third rows, the mobility ratios are 3.77 and 2.16, respectively.

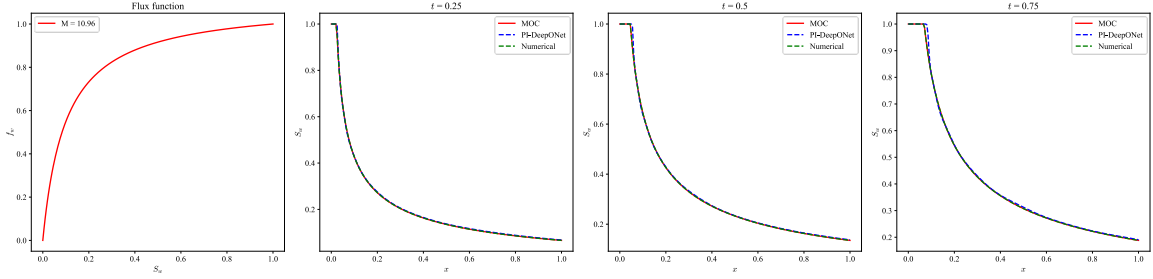

**Fig. 4:** Data-free solution of parametric Buckley-Leverett problem with concave flux functions using PI-DeepONets. The solution profiles three different time steps for a concave flux function generated using a mobility ratio of 10.96.

### 4.2 Nonconvex Flux Function

The interplay between two immiscible fluids flowing through porous media results in highly non-linear relative permeabilities. The Honarpour empirical model is a power-law relationship between a phase's relative permeability and saturation (Equation 5). Using this empirical model with $S_{wc} = S_{or} = 0, K_{rw} = K_{ro} = 1, and\ n_o = n_w = 2$, the flux function reduces to the following quadratic relationship:

$$f_w(S) = \frac{S^2}{S^2 + \frac{(1-S)^2}{M}}. \tag{20}$$

This flux function results in the classical Buckley-Leverett problem as it is known in the literature, and the solution exhibits a shock and a rarefaction wave. Here, we aim to approximate the operator $G_\theta$ mapping the space of nonconvex flux functions to the space of solutions of the Buckley-Leverett equation. We sample 3000 unique mobility ratios from a uniform random distribution such that $1 \leq M \leq 10$. Using these mobilities we generate a training set, $N_{train} = 3000$, of nonconvex flux functions $f^{(i)}_{w_{nonconvex}} = [f_w^{(i)}(s_1), f_w^{(i)}(s_2), \ldots, f_w^{(i)}(s_m)]$ evaluated at $m = 200$ sensor locations (Figure 4), and we take $P = Q = m$. The modified [60] physics-informed DeepONet which consists of branch and trunk networks with two separate 6-layer fully-connected neural networks and 50 neurons per hidden layer is trained for 300,000 iterations using the Adam optimizer with an exponential decay learning rate 0.95 every 2000 steps and, an initial learning rate of $1 \times 10^{-3}$. The loss function (17) is weighted with $\lambda_{IC} = \lambda_{BC} = 5$, and $\lambda_{PDE} = 1$ for an enhanced trainability of the PI-DeepONet.

We note here that Physics-Informed DeepONet also suffers from the same failure mode reported in [38] in the case of physics-informed neural networks when solving hyperbolic transport problems (1) with nonconvex/convex flux functions, see Appendix C. As in [38], a small diffusion term, i.e., second-order derivative term, is added to the right-hand-side of (1) to produce a parabolic PDE in the form of (14). This small diffusion term ($\epsilon \geq 2.5 \times 10^{-3}$) is a scalar diffusion coefficient that represents the inverse of the Peclet number and resembles the numerical diffusion term introduced by discretization using the finite-volume method [38]. The addition of the diffusion term to the right-hand side of the equation allows us to effectively train the physics-informed DeepONet.

Figure 5 shows a comparison between the predicted, the analytical, and the finite difference solutions for three test samples at three different time steps along with their associated flux functions. The mean and standard deviation of the relative $L^2$ error of $N_{test} = 100$ test samples are $3.69 \times 10^{-2} \mp 8.05 \times 10^{-3}$, rspectively. It can be noted from the figures that all solutions exhibit the same structure of a shock followed by a rarefaction wave, albeit the shocks differ in size, speed, and location, depending on the flux function. The behavior of the shocks is very well captured by the trained model. The relative $L^2$ errors for the cases shown in Figure 5 are $3.77 \times 10^{-2}$, $7.58 \times 10^{-3}$, and $1.06 \times 10^{-2}$, for cases a, b, and c, respectively. We note here that the trained PI-DeepONet model achieves more than four orders of magnitude improvement in speed compared to the numerical solver running on the same GPU in inference.

The trained physics-informed DeepONet network displays good accuracy and generalizes well to unseen test functions. However, it does not generalize quite as well as the concave flux functions operator to mobility ratios outside the training range as shown in Figure 6, though the solution is acceptable. In this case, we generate 100 mobility ratios from a uniform random distribution such that $10 \leq M \leq 20$ and generate a new test set. The mean and standard deviation of the relative $L^2$ error of 100 out of range test samples are $9.06 \times 10^{-2}$ and $2.68 \times 10^{-2}$, respectively.

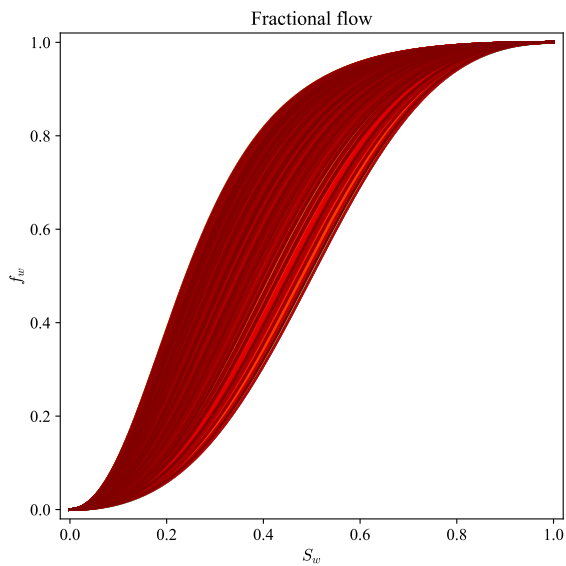

**Fig. 4:** Nonconvex flux functions comprising the training data set ($N_{train} = 5000$).

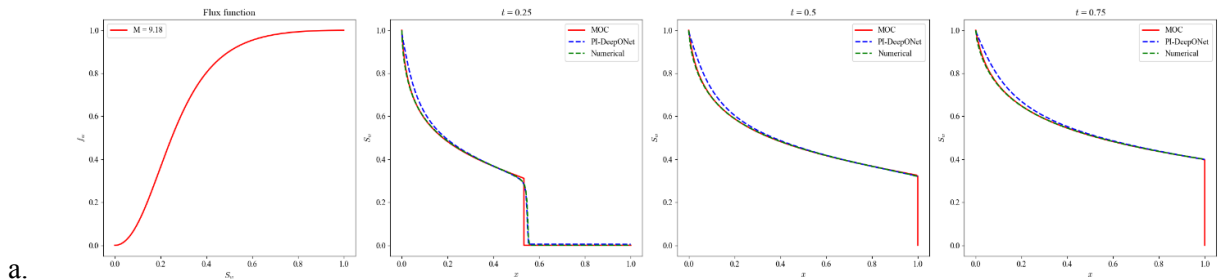

a.

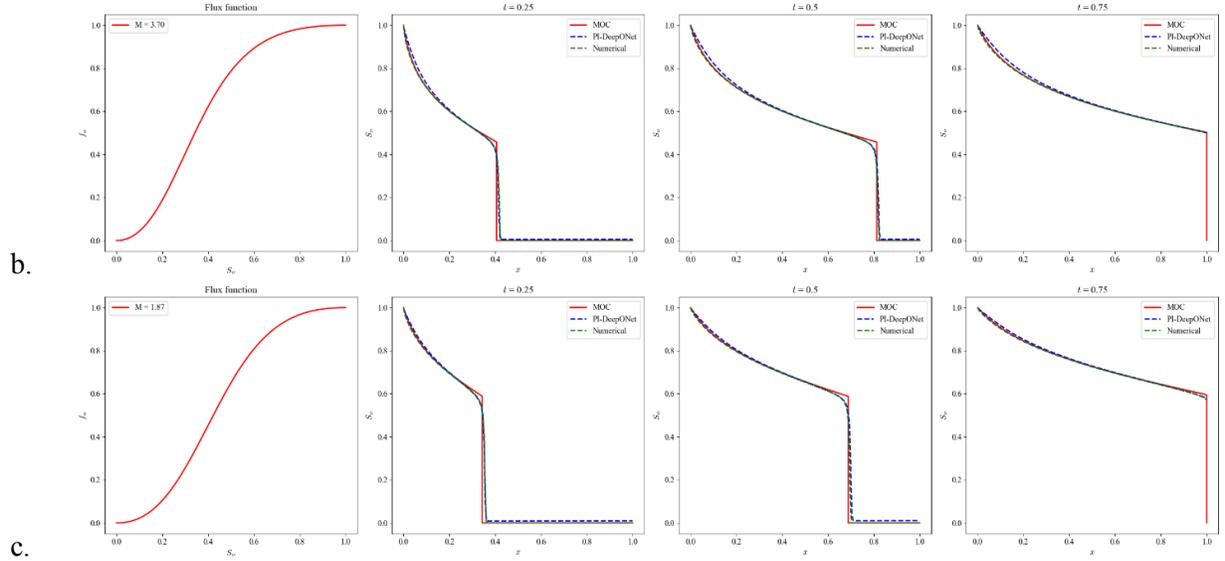

**Fig. 5:** Data-free solution of parametric Buckley-Leverett problem with three different nonconvex flux functions using PI-DeepONets. The first row shows the solution profiles at three different time steps for a nonconvex flux function generated using a mobility ratio of 9.18. In the second and third rows, the mobility ratios are 3.70 and 1.87, respectively.

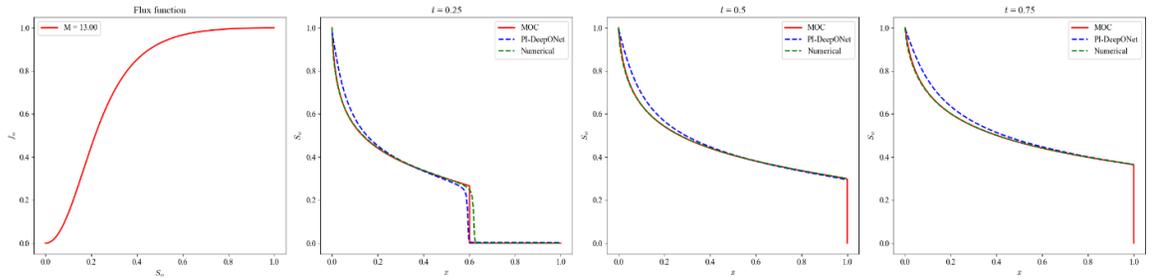

**Fig. 6:** Data-free solution of parametric Buckley-Leverett problem with concave flux functions using PI-DeepONets. The solution profiles at three different time steps for a concave flux function generated using a mobility ratio of 13.0 that is outside the training set parameter space.

### 4.3 Convex Flux Function

The third kind of flux function considered in this study is the convex type, which is simply a quadratic function of saturation $f_w(S) = S^2$. The solution is a self-sharpening wave, propagating as a shock with a unit speed. Consequently, the solution to the Buckley-Leverett problem with a convex flux function is the same irrespective of the mobility ratio. Nonetheless, to train the DeepONet to solve the parametric Buckley-Leverett problem with convex flux functions we generate 1000 (identical) functions to form the training set. We follow this approach regardless of the nonparametric nature of the solution for two reasons; to uphold consistency with the previous test cases, and to evaluate DeepONet's ability to recognize the nonparametric nature of the solution. The standard physics-informed DeepONet architecture is employed, wherein both the branch and the trunk are composed of a 6-layer, fully-connected deep neural networks with 50 neurons per hidden layer. This network is trained over 200,000 iterations using the Adam optimizer, with an initial learning rate of $1 \times 10^{-3}$ that decays by a rate of 0.95 every 2000 steps. The weights in equation (17) are set to $\lambda_{IC} = \lambda_{BC} = 2$, and $\lambda_{PDE} = 1$.

Given the deterministic nature of the solution, we evaluate the trained PI-DeepONet with a single flux function. The relative $L^2$ error is $2.36 \times 10^{-2}$. The solution is shown in Figure 7 at three different time intervals accompanied by its associated flux function. The figure shows that the trained PI-DeepONet model predicts the solutions accurately, unhindered by the nonparametric nature of the problem.

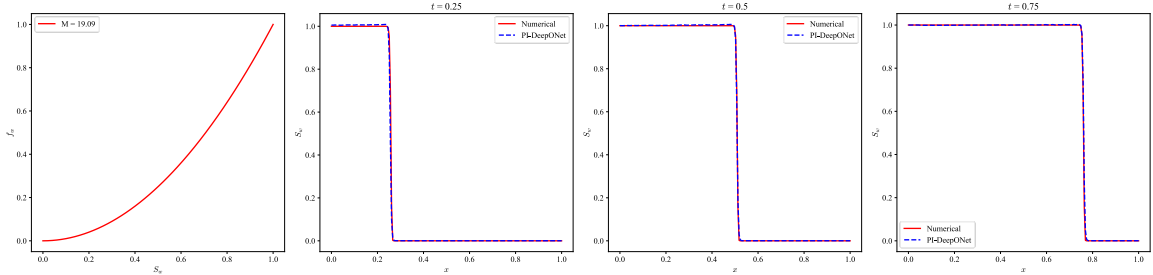

**Fig. 7:** Solution of parametric Buckley-Leverett problem with a convex flux function using PI-DeepONets. The solution profiles at three different time steps for a convex flux function generated using a mobility ratio of 2.65 is shown. This type of flux function has only one solution due to the nonparametric nature of the problem.

## 4.4 Generalized Flux Function

This section is at the heart of our main objective in this work and that is to develop a generic parametric solution for the Buckley-Leverett problem. The solution is based on a generalized flux function that encapsulates all three types of flux functions. This generalized flux function is formulated using relative permeability parameters, which are governed by a set of six variables: viscosities of water and oil, exponents of relative permeability, and endpoints of relative permeability. By considering combinations of these six parameters, we generate a set of relative permeability curves. These curves, in turn, yield flux functions of any of the three types discussed above, which are subsequently employed to learn the parametric solution for the Buckley-Leverett problem.

We train the DeepONet with 5000 unique flux functions (Figure 8) generated using values sampled from a uniform random distribution for the six relative permeability parameters as follows:

- Water viscosities: $0.1 \leq \mu_w \leq 5$
- Oil viscosities: $1 \leq \mu_o \leq 10$
- Water relative permeability endpoints: $0.1 \leq K_{rw}^* \leq 1.0$
- Oil relative permeability endpoints: $0.1 \leq K_{ro}^* \leq 1.0$
- Water relative permeability exponents: $1.0 \leq n_w \leq 3.0$
- Oil relative permeability exponents: $1.0 \leq n_o \leq 3.0$

The generated flux functions in this case can be any of the three types discussed earlier. For instance, for a sample with $n_w = n_o = 1$, the resultant flux function is concave; if $n_w = n_o = 2$, the flux function will be nonconvex, and so on.

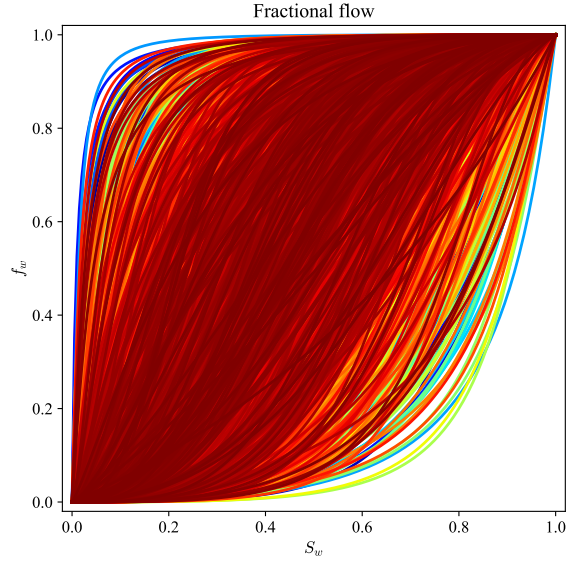

**Fig. 8:** The training set consisting of 5000 flux functions of all three types (concave, convex, and nonconvex).

The modified [60] physics-informed DeepONet which features two separate 8-layer, fully-connected neural networks (branch and trunk networks) with 100 neurons in each hidden layer is trained 400,000 iterations using the Adam optimizer with an initial learning rate of $1 \times 10^{-3}$, decaying every 3000 steps at a rate of 0.95. The loss function (17) is weighted with $\lambda_{IC} = 6$, $\lambda_{BC} = 3$, and $\lambda_{PDE} = 1$.

We then test the trained DeepONet with four new sets of relative permeability (which do not exist in the training set). The four samples are picked in such a way as to display a wide range of behaviors that are captured by the trained DeepONets. The solutions for these test cases are shown in Figure 9 at three different time steps along with their associated relative permeability curves and flux functions. Each of the four figures shows reasonable accuracy and the location of the shock front in Figures 9(a-b) is captured accurately. The small inaccuracies in each of these cases are more than accounted for by the four-order-of-magnitude improvement in speed. The mean and standard deviation of the relative $L^2$ error of $N_{test} = 100$ test samples are $5.83 \times 10^{-2} \mp 5.70 \times 10^{-2}$, respectivly. The relative $L^2$ error for each of the cases presented in Figure 9 are $3.17 \times 10^{-2}$, $1.70 \times 10^{-2}$, $7.76 \times 10^{-3}$, and $7.44 \times 10^{-2}$, for cases a, b, c, and d, respectively.

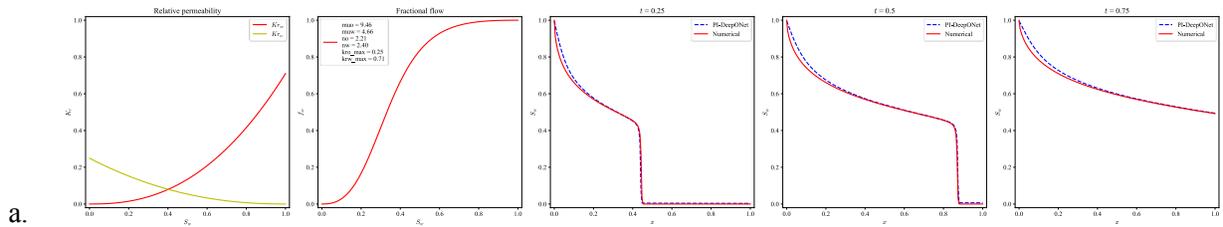

a.

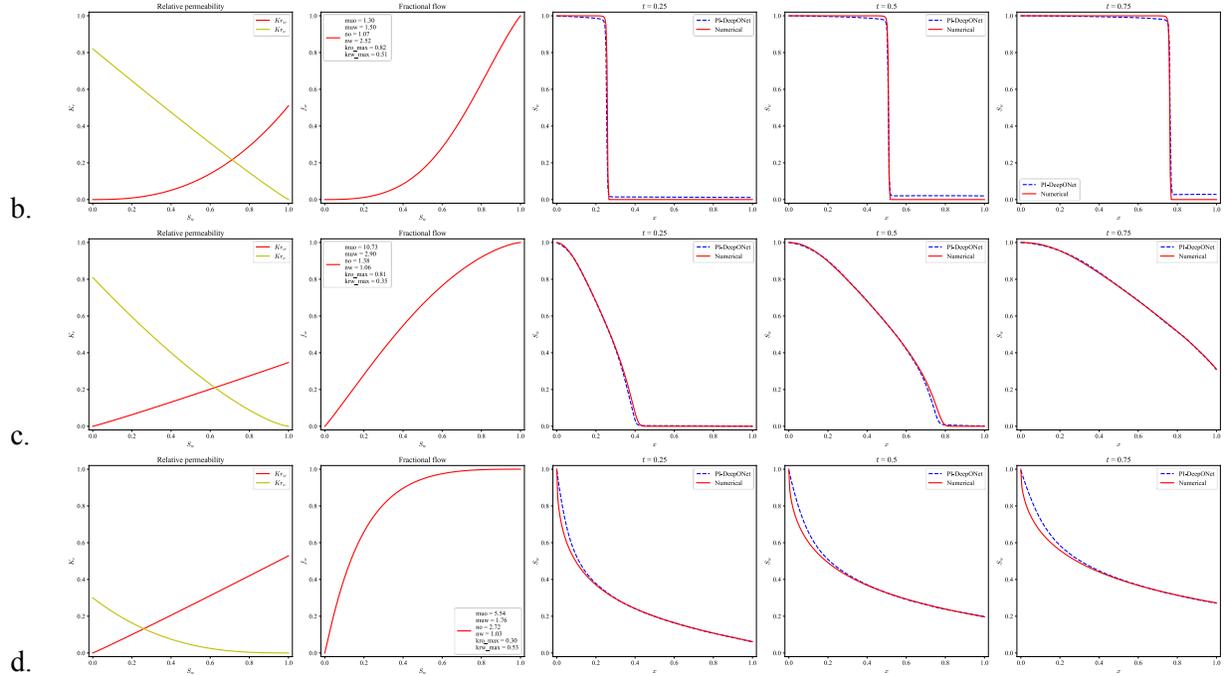

**Fig. 9: S**olution of parametric Buckley-Leverett problem trained on all three types of flux functions concurrently using PI-DeepONets. Each row contains the solution profiles at three different time steps for a unique flux function generated using a set of relative permeability curves shown alongside it; the relative permeability parameters used to generate these flux functions are also shown in the figure.

## 5. Summary, Conclusions, and Future Work

Physics-Informed DeepONets provide an effective and efficient approach to solving two-phase one-dimensional parametric transport problems. The model trained on a wide variety of flux functions resulted in accurate predictions and demonstrated excellent generalization behaviors when evaluated with new flux functions. A summary of the four cases presented in this paper is tabulated in Table 1.

We have shown that our methodology is extremely robust in learning all three types of flux functions simultaneously, resulting in a truly generalized model capable of handling any type of flux function given any set of relative permeability curves and fixed residual saturations (initial and boundary conditions). Moreover, our trained model was able to generalize beyond the parameter space defined by the training set to a certain extent. However, as we move further away from the training parameter space, the model's accuracy begins to decline.

In terms of efficiency, the trained PI-DeepONet model is significantly more efficient in inference (more than four orders of magnitude faster) when compared to traditional numerical solvers running on the same GPU. Here, it is important to keep in mind that the PI-DeepONet model predictions are made over the entire grid ($200 \times 200$) in one shot, while the numerical model solves the problem one-time step at a time. In addition, the numerical model is solved on a ($200 \times 1000$) grid to achieve comparable accuracy.

**Table 1:** Summary of training and testing of the four cases presented in this work.

| Case | Flux function type | $N_{trian}$ | NN size (Depth × Width) × 2 | NN | Iterations × 1000 | Training time (hours) (Nvidia A100) | $N_{test}$ | Average $L_2$ relative error | Approx. Inference time (ms) |
|---|---|---|---|---|---|---|---|---|---|
| 1 | Concave | 1000 | 4 × 100 | FFN | 300 | 0.4 | 100 | $8.53 \times 10^{-3}$ $\mp 1.84 \times 10^{-3}$ | 0.72 |
| 2 | Nonconvex | 3000 | 6 × 50 | Modified | 300 | 1 | | $3.69 \times 10^{-2}$ $\mp 8.05 \times 10^{-3}$ | 0.86 |
| 3 | Convex | 1000 | 6 × 50 | FFN | 200 | 0.5 | | $2.36 \times 10^{-2}$ $\mp 0.0$ | 0.79 |
| 4 | Any | 5000 | 8 × 100 | Modified | 400 | 2.5 | | $5.083 \times 10^{-2}$ $\mp 5.70 \times 10^{-2}$ | 0.84 |

Lastly, a logical question that arises is whether the DeepONet architecture is the optimal choice for the parametric transport equation. DeepONet employs feed-forward neural networks to extract feature vectors from the input functions and spatiotemporal coordinates. However, there's scope for alternative neural networks to be used instead of feed-forward neural networks (FFN), such as a convolutional neural network or a residual neural network [63]. Convolutional neural networks may be particularly advantageous for equispaced domains [11]. Moreover, innovative advances in this field propose fundamentally different approaches. For instance, the physics-informed neural operator (PINO) [64], which utilizes the Fourier Neural Operator (FNO) architecture, learns the parameters of the integral kernel in Fourier space. Similarly, Physics informed Wavelet Neural Operator (WNO) [15] carries out network parameterization in the Wavelet space. These methods highlight the diverse range of approaches that could potentially be explored. A comprehensive and systematic study, however, would be necessary to determine the most effective approach. More importantly, these networks, inherently exhibit strong potential for applications in stochastic modeling and uncertainty quantification, considering that they address parametric PDEs, as opposed to deterministic PDEs. Addressing these application areas will truly unlock the immense potential of machine learning for flow and transport in porous media.

## Appendix A   Modified DeepONet Architecture

In [60] a modified DeepONet architecture was proposed to improve the trainability and predictive accuracy of the PI-DeepONet. In this work, the modified DeepONet architecture is used in cases 4.2 and 4.4 demonstrating considerable improvement in the overall results. This particular architecture features two encoders applied to the inputs of the trunk and branch networks, which enhances the networks' ability to handle nonlinearities. For completeness, the forward pass of the modified neural network is as follows:

$$\boldsymbol{U} = \sigma(\boldsymbol{X}\boldsymbol{W}_1 + \boldsymbol{b}_1), \qquad \boldsymbol{V} = (\boldsymbol{X}\boldsymbol{W}_2 + \boldsymbol{b}_2),$$

$$\boldsymbol{H}^{(1)} = \sigma(\boldsymbol{X}\boldsymbol{W}^{(l)} + \boldsymbol{b}^{(l)}),$$

$$\boldsymbol{Z}^{(k)} = \sigma(\boldsymbol{H}^{(k)}\boldsymbol{W}^{(z,k)} + \boldsymbol{b}^{(z,k)}), \qquad k = 1, \dots, L$$

$$\boldsymbol{H}^{(k+1)} = (1 - \boldsymbol{Z}^{(k)}) \odot \boldsymbol{U} + \boldsymbol{Z}^{(k)} \odot \boldsymbol{V}, \qquad k = 1, \dots, L \qquad (21)$$

$$\boldsymbol{f}_\theta(X) = \boldsymbol{H}^{(L+1)}\boldsymbol{W} + \boldsymbol{b},$$

$$\theta = \{\boldsymbol{W}^1, \boldsymbol{b}^1, \boldsymbol{W}^2, \boldsymbol{b}^2, (\boldsymbol{W}^{z,l}, \boldsymbol{b}^{z,l})_{l=1}^L, \boldsymbol{W}, \boldsymbol{b}\}.$$

The parameters of this architecture are the same as the MLP, and $\odot$ denotes element-wise multiplication. $\boldsymbol{U}$ and $\boldsymbol{V}$ are two transformer networks that project the input variables to a high-dimensional feature space.

## Appendix B  Grid Sensitivity

The numerical solution scheme used here to solve the Buckley-Leverett problem is the explicit upwind finite difference method which is a common method for solving hyperbolic partial differential equations [34]. The upwind scheme takes the following form:

$$u_i^{n+1} = u_i^n - \frac{\Delta t}{\Delta x}[f(u_i^n) - f(u_{i-1}^n)].$$

The two-point scheme uses a one-sided difference in the direction where the wave is coming (boundary condition). The key features of this method are its simplicity and its ability to handle advection-dominated flow. This approach is first-order accurate and much less diffusive than other methods, which makes it ideal for solving hyperbolic transport problems of the Riemann type. However, it imposes stringent restrictions on the time-step (known as the CFL number):

$$\max_{i,n}|f_c'(\cdot\,;u_i^n,u_{i+1}^n)|\frac{\Delta t^n}{\Delta x_i} \leq 1.$$

As a result, we use a constant $\Delta t = 0.001$ and vary the $\Delta x$ between 0.1 and 0.0025 to always abide by this condition and to obtain an accurate solution. Figure 10 shows the solution of the hyperbolic Buckley-Leverett problem with a nonconvex flux function, where $M = 2$ at $t = 0.25$ for different $\Delta x$'s. As can be seen in the figure, a $n_x \sim 400$ results in the most accurate solution that mimics the analytical solution, and produces correct approximations of both the leading shock wave and the trailing rarefaction wave.

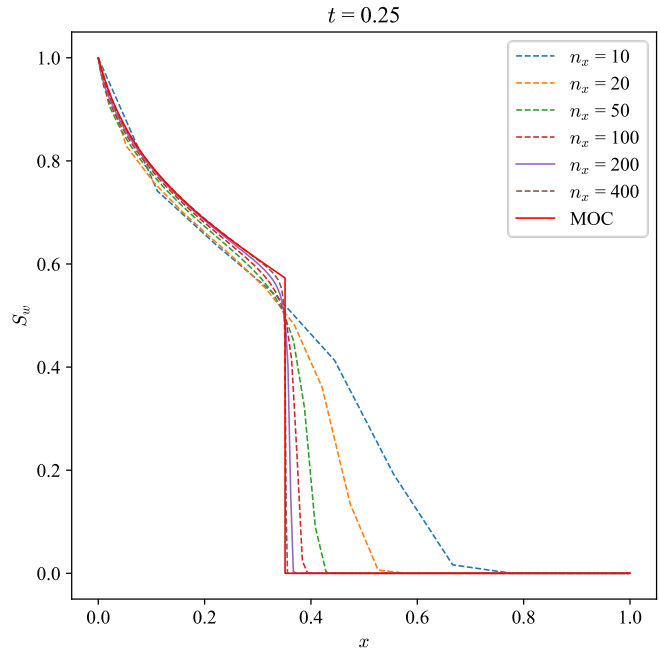

**Fig. 10:** Grid sensitivity study on the effect of the spacial grid size on the accuracy of the solution with a constant $\Delta t = 0.001$. The solution of the Buckley-Leverett problem with a nonconvex flux function at $t = 0.25$ is shown for $\Delta x = 0.1, 0.05, 0.02, 0.01, 0.005,$ and $0.0025$ along with the analytical solution using the method of characteristics (MOC).

## Appendix C   Nonconvex without Diffusion Term

Physics-Informed DeepONet suffers from the same failure mode reported in [38] in the case of physics-informed neural networks when solving hyperbolic transport problems with nonconvex/convex flux functions.

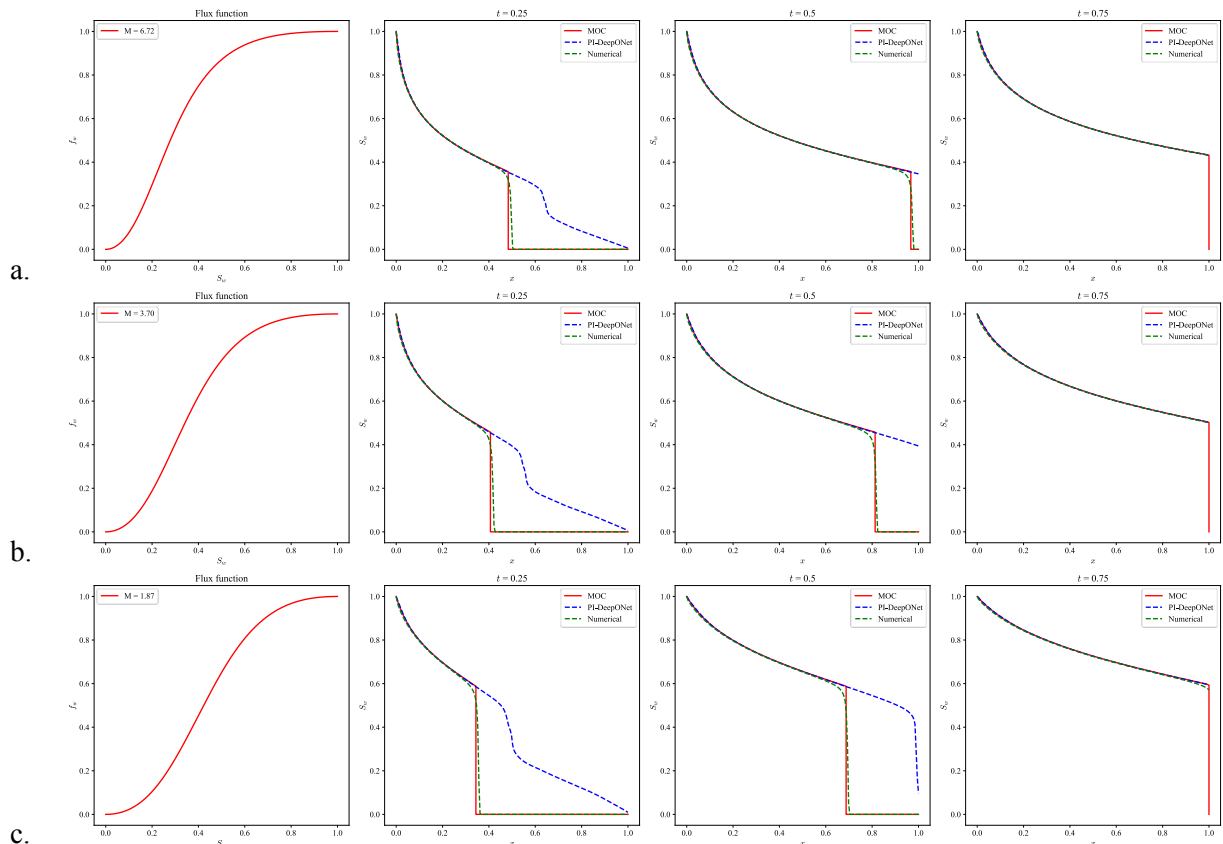

**Fig. 11:** Failed solution of parametric Buckley-Leverett problem with three different nonconvex flux functions using PI-DeepONets without a diffusion term in the residual. The first row shows the solution profiles at three different time steps for a nonconvex flux function generated using a mobility ratio of 9.18. In the second and third rows, the mobility ratios are 3.70 and 1.87, respectively.